\documentclass{elsarticle}
\usepackage{amsmath}
\usepackage{amsthm}
\usepackage{amssymb}
\usepackage{enumerate}
\usepackage{verbatim}
\usepackage{graphics}
\usepackage{tikz}
\usetikzlibrary{shapes}
\newcommand{\junk}[1]{}
\def\eredmeny{O\left ( {n}m^{0.585} \right)}

\newtheorem{theorem}{Theorem}

\theoremstyle{definition}

\begin{document}
\begin{frontmatter}
\title{Modulated String Searching\tnoteref{gen}}

\author[axa] {Alberto Apostolico\fnref{axa1}}
\author[renyi]{P\'eter L. Erd\H os\fnref{elp}}
\author[renyi]{Istv\'an Mikl\'os\fnref{mik}}
\author[uea]{Johannes Siemons}
\address[axa]{ College of Computing, Georgia Institute of Technology, \\
        801 Atlantic Drive,  Atlanta, GA 30318, USA\\
        {\em and} \\
       Istituto di Analisi dei Sistemi e Informatica, \\
       Consiglio Nazionale delle Ricerche, Viale Manzoni 30, Roma, Italy, \\
       {\tt email}: axa@cc.gatech.edu}
\address[renyi]{Alfr\'ed R{\'e}nyi Institute of Mathematics, Re\'altanoda u 13-15 Budapest, 1053 Hungary\\
        {\tt email}: $<$erdos.peter,miklos.istvan$>$@renyi.mta.hu}
\address[uea]{School of Mathematics, University of East Anglia, Norwich, UK, \\
       {\tt email}: j.siemons@uea.ac.uk}
\fntext[axa1]{Additional support was provided by the United States-Israel Binational
    Science Foundation (BSF) Grant No. 2008217 and by the Research Program of Georgia Tech.}
\fntext[elp]{Research supported in part by the Hungarian NSF under contract NK 78439
    and  K 68262.}
\fntext[mik]{Research supported in part by the Hungarian NSF under contract PD 84297.}
\tnotetext[gen]{This research was carried out in part while A. Apostolico and J. Siemons were visiting   the R\'enyi Institute, with support from the Hungarian Bioinformatics MTKD-CT-2006-042794 and Marie Curie Host Fellowships for Transfer of Knowledge. }

\begin{abstract}
\noindent
In his 1987 paper entitled   {\sl Generalized  String Matching} Abrahamson introduced  the concept of {\em pattern matching with character classes} and provided the first efficient algorithm to solve this problem. The best known solution to date is due to Linhart and Shamir (2009).

Another broad yet comparatively less intensively studied class of string matching problems is numerical string searching, such as for instance "less-than" or $L_1$-norm string searching. The best known solutions for problems in this class are based on FFT convolution after some suitable re-encoding.

The present paper introduces  {\em modulated string searching} as a unified framework for string matching problems where the numerical conditions can be combined with some Boolean/numerical decision conditions on the character classes. One example problem in this class is the {\em locally bounded $L_1$-norm} matching problem with parameters $b$ and $\tau:$  here the pattern "matches" a text of same length if their $L_1$-distance is at most $b$ and if furthermore there is no position where the text element and pattern element differ by more than the local bound $\tau.$ A more general setup is  that where the pattern positions contain character classes and/or each position has its own private local bound.  While the first variant can clearly  be handled by adaptation of the classic FFT method, the second one is far too complicated for this treatment.  The algorithm we propose  in this paper can solve all such problems efficiently.

The proposed framework contains two nested procedures. The first one, based on Karatsuba's fast multiplication algorithm, solves pattern matching with character classes within time $O\left ( {n} m^{0.585} \right)$, where $n$ and $m$ are the text and pattern length respectively (under some reasonable conventions).  This is slightly better than the complexity of Abrahamson's algorithm for generalized string matching but worse than algorithms based on  FFT. The second procedure, which works as a plug-in within the first one and is tailored to the specific problem variant at hand, solves the numerical and/or Boolean matching problem with high efficiency. Some of the previously known constructions can be adapted to match or outperform several (but not all) problem variations handled by the construction proposed here. The latter aims to be a general tool that provides a  unified  solution for all problems of this kind.\end{abstract}
\end{frontmatter}

\section{Introduction}\label{sc:intro}
\noindent String searching is a basic primitive of computation. In the standard formulation of the problem, we are given a pattern and a text and are required to find all occurrences of the pattern in the text. Several variants of the problem have also been considered, such as allowing mismatches, insertions, deletions, swaps and so on.

In his paper~\cite{abra} Abrahamson introduced the notion of {\em pattern matching with character classes\,} (or {\bf PMCC} for short)  which is specified as follows. The {\em pattern} $P$ of length $m$ is given as a sequence of character
\junk{\marginpar{\scriptsize\color{bl} From J: Why is pattern in italics while texts is not. I am not sure about the difference between `', ``'' and italics. It should be consistent. }}
classes ($P[j]\subseteq \Sigma$) and the {\em text} $T$ is a sequence from $\Sigma^*$ (that is $T[i] \in \Sigma)$. Here $P$ occurs at location $i$ in $T$ if $\forall j: 1\le j \le m, \ T[i+j-1] \in P[j].$
The problem of PMCC for a (typically long) text is to find all positions in the text $T$ where
the pattern $P$ occurs.  Standard string searching thus corresponds to the special case where each character class consists of exactly one element. In the original formulation PMCC was called {\em generalized pattern matching}.

Abrahamson proved that PMCC is harder than standard string searching and gave an algorithm for it.  Since the algorithm deals with unrestricted alphabets, both the text and the pattern are encoded over a fixed auxiliary alphabet. Now, let $\hat M$ denote the number of symbols used over the original alphabet to describe the pattern elements, and let $ M $ be the total length of the encoding of the pattern. Likewise, let $n$ be the number of symbols in the text sequence, and $N$ the total length of the encoding of the text. Then the time complexity of Abrahamson's algorithm is
$$
O\left (M+N+ n \hat M^{1/2} \mathrm{polylog}(m)\right ).
$$
The state of the art for PMCC is due to Linhart and Shamir~\cite{LS}. Their algorithm has the following  impressive time complexity: having set  $\kappa = \log_{|\Sigma|} (\log n/ \log m)$, then it is $O(n \log m )$ for $\kappa\le 1$, while for  $\kappa>1$ it becomes  $O(n \log (m/\kappa)).$ Their approach can be extended to solve PMCC with mismatches and to PMCC with subset matching. It is based on encoding the text and pattern using large prime numbers, and on an FFT-based convolution process. It is suitable for checking "element(s) in a subset relation" but not for more complicated conditions.

The problem of searching for strings consisting of numerical values rather than characters arises in countless applications and some variants have  already been studied in combinatorial pattern matching.  In these problems the {\em fitting\,} conditions are described  in numerical terms. For example, in the  {\em less-than} string searching  problem (Amir and Farach \cite{AF}), the pattern fits the text if at each positionof the alignment the pattern value does not  exceed the corresponding text value. Additional variants require the computation of the {\em $L_1$-distance} of the pattern from the text at each starting position (Amir, Landau and Vishkin~\cite{Amir},  Lipsky~\cite{Lipsky}). Yet another version, known as the {\em $k-L_1$-distance problem\,}  (Amir, Lipsky,  Porat and Umanski~\cite{ALP05}), consists of computing approximate matching in the $L_1$-metric.

These fast methods are also based on suitable encoding processes and on FFT, with corresponding time complexity. These algorithms do not seem to be applicable to numerical string searching with character classes and in general to those cases where a pointwise evaluation of individual comparisons is required.

\medskip\noindent
In the next section we introduce the {\em modulated string searching\,} framework (or {\bf MSS} for short) which combines the flexibility of PMCC with numerical calculations and/or more complicated Boolean conditions. We will give first a simple and na\"\i ve solution for the problem. (See Section \ref{sc:MSS}.)

Our proposed approach for MSS is by a pair of nested procedures. The first one (see Section \ref{sec:approx}) is an algorithm to solve PMCC, based on Karatsuba's fast multiplication  method.  Its  complexity  is
$$
\eredmeny
$$
where $n$ and $m$ are the text and pattern lengths, respectively, provided that all other parameters involved such as character class number etc. can be  treated  as constants.
\junk{\marginpar{\scriptsize\color{bl} J: I dont understand the sentence (in cases...), something is missing, Generally we should avoid brackets, it makes things less clear}
(in cases when the number of the different subsets occurring in the pattern together the "private" bounds can be considered constant). }  Here one could argue that the application of the Toom-Cook or the  Sch\"onhage algorithms~\cite{CP, Sch} yields a better performance. This is  true, however, only for certain values of the text and pattern lengths. In addition, those algorithms require  higher overheads, offsetting the overall gain.
The above complexity is also worse than the complexity achieved in, say,  \cite{LS}. However, the present method allows us to design a second procedure which works as a plug-in within the first one (see Section \ref{sc:numera}) and which solves a variety of numerical and/or Boolean problems. Indeed, the first procedure of our framework is always the same, while the plug-in procedure and its complexity depend heavily on specific matching conventions.
Some of the previously known constructions can be adapted to match or outperform several (but not all)  problem variations handled by the method  proposed here, which therefore aims to be a general tool that provides a  unified  solution for all problems of this kind.

\section{Modulated string searching framework}\label{sc:MSS}
\noindent
The framework for {\em modulated string matching  on character classes} is as follows. The alphabet $\Sigma$ is some set of  natural numbers and $b$ denotes an absolute constant. \marginpar{\scriptsize
\junk{\color{bl} J: What is $m?$  Is this what is meant?}}
The pattern is a string of character classes (each class being a finite subset of $\Sigma$) whose length, that is the total number of the character classes,  is denoted by $m$ and the text is a finite string over $\Sigma$. The matching conditions are dictated by two functions with the following features. One of them depends on the particular variant of the problem and takes as arguments a character class and a character, and returns in constant time a score of the match.  The second function takes as arguments the scores at the $m$ positions of an alignment of the pattern against the text and returns true in case they add up to at most $b$, false otherwise.

\smallskip\noindent {\bf Examples:} Consider first the locally bounded $L_1$-distance string matching problems on character classes: Assume we are given two strings of equal length $m$ over the natural numbers.  Then the  {\em $L_1$-distance} of these strings  is  $\,\sum_{i=1}^m |P_i -T_i|,$ as usual. When one of the strings is given with character classes, the {\em $L_1$-distance on character classes}  at a given position is defined as the smallest $L_1$-distance between the element of the first string at that position and any of the elements in the effacing class. For given pattern and text strings, the total distance at some starting text position is the sum of the above local distances. Let now $b$ and $\tau$ be absolute constants. We say that the pattern {\em fits} at a given position of the text  in {\em locally bounded $L_1$-distance} with parameters $b$ and $\tau$  if there is no position in the corresponding substring of the text where the $L_1$-distance from the pattern element is bigger than $\tau$ and, in addition,  the total $L_1$-distance of the two strings is at most $b$. The {\em locally bounded $L_1$-distance string matching problem on character classes} then is to find all positions of the text where the pattern fits.

When each pattern class consists of only one element then one can easily design a two-phase FFT based algorithm to solve  this problem efficiently. However, if the classes are not singletons and / or each pattern position has its own private local bound then this is not feasible anymore.

A closely related notion is the {\em $\tau$-truncated $L_1$-distance}. For two strings of length $m$ this parameter is defined as $ \sum_{i=1}^m \min (|P_i -T_i|, \tau) $. This can be visualized as testing a sequence of manufactured  items against a standard of reference: the difference at each  position describes, e.g.,  the cost to repair a token {\it in situ} while it is more economical  to replace that token when the repair becomes too costly. When the pattern is given in form of  character sets then the obvious analogous definition applies. Then the {\em $\tau$-truncated $L_1$-distance string matching problem  on character classes} becomes to find all positions of the text where the $\tau$-truncated $L_1$-distance of the pattern is at most $b.$  In particular, in case of singleton character classes in the pattern and a big enough constant $\tau$ this yields the standard $L_1$-distance problem.

We finally list a  third example which can be called {\em fitting assignment}. Here we simply have a map $\mathcal{A}$ from the pairs of text characters and pattern character classes (which may each specify its position within the pattern), say, to the integers $\mathbb{Z}$, or even more generally, to the reals $\mathbb{R}.$ In this case we have to store the values for all possible pairs, but, in exchange, there is no need to calculate anything. It is clear  that no FFT based method can solve this problem. On the other hand, these and other problems can be described and solved in the framework proposed here.

\medskip\noindent
Modulated string searching can be solved easily by the following direct method. Align the pattern with the text starting at every position of the text. Each text character is matched against its corresponding set $P_i$. In most cases (like in the first two described above) finding out whether a text character fits into $P_i$ can be managed with the help of a simple merge operation and so  requires roughly $ \log |P_i|$ time. The algorithm then compares the distance of the text element and the neighboring pattern elements against  the threshold $\tau$. Adding up for all text characters this yields $n \sum_{i=1}^m  \log \big | P_i \big |$  time.

\bigskip\noindent
Before proceeding further we specify a more convenient representation of the pattern elements. For each pattern position $i$ we have in general a character class $P_i$ and we will represent this subset of the alphabet $\Sigma$ by a binary  characteristic vector $\bar p_i$ of length $|\Sigma|$. Since there are several kinds of\, "length" \,in this paper, we will use the term {\em dimension\,} for the length of a vector. So $\bar p_i$ is a vector of dimension $|\Sigma|.$ If we represent our text symbols analogously by characteristic vectors  $\bar t_j$ for all $j=1,\ldots,n,$ each one of which contains exactly one non-zero element, then the text character $T_j$ and the pattern character class $P_i$  match if and only if the scalar product $\langle \bar p_i, \bar t_j \rangle $ of the corresponding characteristic vectors is exactly $1.$

With this notation, for each $j=0,\ldots, n-m-1,$ the substring of $T$ starting at position $j+1$ and ending at $j+m$ fits the pattern string  if and only if
\begin{equation}\label{eq:scalar}
v_{j+1}:=\sum _{i=1}^m  \langle \bar p_i, \bar t_{j+i} \rangle
\end{equation}
equals $m$ exactly. Furthermore, when $v_{j+1}= m- \ell,$ then we have exactly $\ell$  mismatches.

The direct computation of the above sums would require $O(nm)$ scalar products where each one may take $O(|\Sigma|)$ time to compute. In the next section we show how one can speed up this algorithm for the MSS problem using a  convolution-type argument.

\section{PMCC with Karatsuba's fast multiplication algorithm}\label{sec:approx}
\noindent In this section we develop an algorithm to solve the PMCC problem based on Karatsuba's fast multiplication (Karatsuba and  Ofman \cite{KO1963}). Recall that Karatsuba's algorithm requires
$$O \left ( m^{\log_2 3} \right ) $$
single digit multiplications and as many additions to multiply two polynomials of degree $m-1.$
Now if we want to multiply two polynomials $f$ and $g$ of degrees  $n$ and $m = n/q$ respectively we first split $f$ into  segments $f_1,\ldots, f_q$ of length $m,$ then  carry out all multiplications $f_i g$ and finally add up the results using the corresponding place values for all results. In other words, we compute
$$
f \cdot g= \sum_{i=0}^{q-1} (f_i \cdot g) x^{im}\,.
$$
We shall see  (Expression \ref{lm:chunk-comp})  that for
a constant  number of different character classes this requires  altogether  roughly $\eredmeny$ time.

In conclusion, we carry out  $q=n/m$ polynomial multiplications and suitable combine the results into our final answer. Therefore, at the heart of our application we face the following problem:

We are given two strings of equal length $m$ consisting of binary vectors of dimension $|\Sigma|$ where each vector in the first sequence has exactly one non-zero element. We want to compute the "product" of the two strings in such a way that for each position $j$ the result is exactly $v_j$ as defined above.  Note that this actually corresponds to solving an extension of exact search, since the $v_j$'s now yield as a byproduct also
the number of possible mismatches in correspondence with each alignment. For the simplicity of this discussion it is convenient to assume that the length of the strings is a power of 2. This does not affect generality since any string can be padded suitably with zeroes.  We remark in passing that
we could extend our approach by allowing character classes in the text as well (that is, several \,$1$s\, in the corresponding characteristic vectors).  Leaving such a generalization for an exercise,
we will  now establish the following fact:
\begin{theorem}
Under suitable bounds on the number of distinct character classes, numerical values and position-specific thresholds, the problem of modulated string searching (with possible mismatches) can be solved by an adaptation of Karatsuba's multiplication algorithm  in time $\eredmeny.$
\end{theorem}
\junk{
\noindent{\color{bl} \scriptsize From J: I dont like definition inside a theorem, and I find the inverted commas and ``''  too vague. Instead we could define a class ${\cal C}$ of problems, also elsewhere we have been speaking of classes of problems. The theorem then says that all problems in class ${\cal C}$ can be dealt with in time  $\eredmeny$ by an adaptation of Karatsuba's multiplication algorithm. }
    }

\medskip
\noindent
A detailed analysis of complexity will follow after Equation (\ref{lm:chunk-comp}).

The proof  requires us to revisit  Karatsuba's algorithm carefully. Recall that this algorithm is based on the following trick originally invented by Gauss for multiplying complex numbers:
Let $f, g $ be two polynomials of degree $2k - 1,$ let $a,b,c$ and $d$ be polynomials of degree $ k-1$ and
set $f = a x^{k} +b$ and $g=c x^{k}+d$. Then
\begin{equation}\label{eq:kara}
(a x^{k} + b) (c x^{k}+d)= a c\cdot  x^{2k}+ [(a+b)(c+d) -ac-bd] \cdot x^{k}  + b d
\end{equation}
The algorithm computes all products recursively.  Figure 1 displays the basic recursion. Its control structure borrowed from the pseudocode in  Weimerskirch and  Paar~\cite{WP2003} is reproduced  here for the convenience of the reader.

\bigskip {\tt
\begin{center}
          \begin{tabular}{l l}
        &      {\bf Algorithm KAM } ~~ $z = KAM(f,g)$ \\
        &      {\bf Input:} Polynomials $f(x), g(x)$; ~~$2k = degree(f) +1  = degree(g) +1 $.\\
        &      {\bf Output: } $z(x)  = f(x) \times g(x) $  \\
        &      {\bf if } $2k = 1$ {\bf return} $f \times g$ \\
        &      {\bf set } $f(x) = a(x) x^{k} +b(x)$; ~~ $g(x)=c(x)x^{k}+d(x)$\\
        &      {\bf create} $r_1(x) = a(x)+ b(x), ~~r_2(x) = c(x)+ d(x) $ \\
        &      $t_1 \leftarrow $ {\bf KAM}$(a,c)$\\
        &      $t_2 \leftarrow $ {\bf KAM}$(b,d)$\\
        &      $t_3 \leftarrow $ {\bf KAM}$(r_1, r_2)$\\
        &      {\bf return} $t_1x^{2k} + (t_3 - t_1 - t_2)x^{k} + t_2$\\
          \end{tabular}
\end{center}
	}
\noindent
Thus, for suitable constants $\gamma$ and  $\delta$ the number of elementary operations performed by the algorithm is governed by the recurrence
\begin{equation}
T(2k) =  3 T( k ) + 2\gamma k +\delta
\end{equation}
for which the Master Theorem gives the asymptotic bound $T(k) = \Theta (k^{\log_23}).$ More specifically, the recursive procedure requires altogether $k^{\log_2 3}$ multiplications and not more than $ 6k^{\log_2 3} - 8k + 2$ elementary additions and subtractions (see \cite{WP2003}).

\medskip
\begin{figure}[h]\caption{The structure of Karatsuba's algorithm.}
\begin{center}
\begin{tikzpicture}
\tikzstyle{level 1}=[sibling distance=4cm]
\tikzstyle{level 2}=[sibling distance=4cm]
\tikzstyle{level 3}=[sibling distance=1cm]
\node (root) [ellipse,draw,very thick,sibling distance=4cm] {$(a x^k + b) (c x^k +d)$}
    child [very thick]{node[ellipse,draw,very thick,sibling distance=4cm] {$ac x^{2k}$}
        child [very thick]{node[ellipse,draw,very thick] {$a_1c_1 x^{2k}$}}
        child [very thick] {node[ellipse,draw,very thick] {$(a_1+a_2)(c_1+c_2)x^k$}
        child [very thick]
        child [very thick]
        child [very thick]}
    child [very thick]{node[ellipse,draw,very thick] {$a_2 c_2$}
        child [very thick]
        child [very thick]
        child [very thick]}
        }
    child [very thick]{node[ellipse,draw,very thick] {$(a+b)(c+d)x^k$}}
    child [very thick]{node[ellipse,draw,very thick] {$bd$}
       child [grow=-120,very thick]
       child [grow=-90,very thick]
       child [grow=-60,very thick]}
;
\end{tikzpicture}
     \label{fig1}
\end{center}
\end{figure}
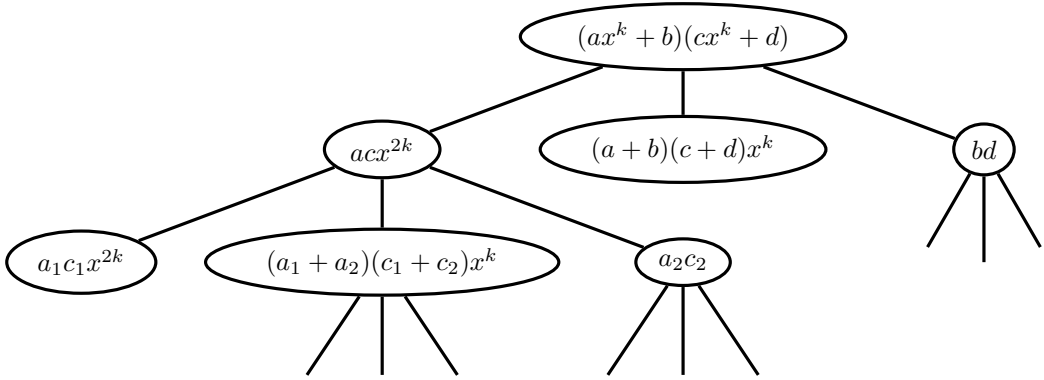

\noindent  The main idea behind our proposed nested procedure is to substitute the regular multiplications (over the underlying number domain) performed in the leaves of our recursion tree with some symbolic computation over a not necessarily commutative ring structure $\mathcal{R}$ over the ring $\mathbb{Z}$ of integers.  Strictly speaking, this symbolic computation constitutes our plug-in procedure and provides the flexibility of our proposed approach. However, once the Karatsuba algorithm is extended in this way to work on the polynomials in $\mathcal{R}[x]$,  the complexity analysis above does no longer apply automatically.  We thus need  to take a closer look at the original algorithm's mechanics and complexity.

Observe that {\em before\,} issuing the recursive calls, the procedure needs only to perform two additions of polynomials. When control {\em returns\,} from the recursion, it needs to perform 4 such operations (2 additions and 2 subtractions), whence a total of 6. The mechanism of the original Karatsuba algorithm within the recursive calls can be subdivided into three phases:
\begin{enumerate}[{\rm {Phase} 1}]
\item Proceeding top-down,  the procedure computes the items at each branching point and (finally) at the leaves of the recursion tree. At each of the $3^h$ vertices at level $h$ it performs additions of  two  polynomials of length $k/{2^h}$, each requiring $k/{2^h}$ elementary additions over the ring $\mathcal{R}$. So the actual time complexity of these operations in our generalized Karatsuba algorithm will depend on the complexity of the coefficients from $\mathcal{R}$ and their representations.
\item At each of the $k^{\log_2 3}$  leaves, the procedure performs the required pairwise product between monomial coefficients. (Again this cost depends on the actual formal ring.)
\item Proceeding now bottom-up, the rocedure computes the actual polynomial values in correspondence with each branching vertex. This is done by shifting the values of the three children by $0$, $\ell$ and $2\ell$ positions accordingly, and performing two addition and two subtraction polynomials of length $4\ell$.
\end{enumerate}

\noindent
In Phase 1 the original algorithm performs roughly $2 k^{\log_2 3}$ elementary additions (over the underlying number domain) while in Phase 2 there are   $ k^{\log_2 3}$ elementary pairwise products and additions. Finally, in Phase 3 roughly $4  k^{\log_2 3}$ elementary additions of $\mathcal{R}[x]$-elements are performed.

\medskip\noindent
We give next a formal description of the objects of our algorithm:  Let the set $\Omega$ consist of one symbol for each subset of $\Sigma$ that appears anywhere in the pattern. Furthermore let $\Gamma:=\Sigma \cup \Omega.$ In what follows, the generic elements of these sets are denoted by $\gamma, \sigma$ and $\omega$ and we adopt these symbols as the appropriate characteristic vectors of the subsets. When, like in the case of locally bounded $L_1$-distance problem, the pattern positions contain their private local bounds as well, then we have  $\omega$'s representing the same subset but containing different private local bounds.
Next we consider the free ring  $\mathcal{R} =\mathbb{Z}\Gamma$ with generators $\Gamma$ over the integers\footnote{As  is well known,  the elements of this ring are formal linear combinations of finite words over $\Gamma$ with coefficients in $\mathbb{Z}$. The additive group is commutative. The multiplication of two finite words is the concatenation of the words and hence is non-commutative. This multiplication extends distributively to the ring. The multiplication by scalars is distributive.  }.
\junk{
\marginpar{\scriptsize \color{bl} From J: as notation I would prefer $\mathbb{Z}\Gamma^{*},$ as we have words in $\Gamma$ rather than single lmembers from $\Gamma.$ If we accept this then notation needs to be changed later.} }

Recall that we are representing any pair formed by the pattern $P$ and a corresponding segment $T'$ of length $m$ from our text $T$  by the polynomials
\begin{equation}\label{eq:poly}
T'(x) \in \mathbb{Z}\Gamma [X], \qquad T'(x)= \sum_{i=1}^m \mu_i x^{m-i},
\end{equation}
and
\begin{equation}\label{eq:poly1}
P(x) \in \mathbb{Z}\Gamma [X], \qquad P(x)= \sum_{i=1}^m \nu_i x^{i-1};
\end{equation}
where each $\mu_i \in \Sigma$ and $\nu_i\in \Omega.$ In Phase 1, in the "middle" child of each vertex,  we need to evaluate the sum of two polynomials. The coefficients of these polynomials are elements of the free ring, that is, they are formal linear combinations of the generators
%(and no concatenations of more than one generator),
since, except at the leaves, along the algorithm no (symbolic) multiplication takes place among ring elements. Furthermore  these combinations never consist of more than $|\Sigma|$ generators in the left polynomials (coming from the text) and more than $|\Omega|$ generators in the right polynomials (coming from the pattern).

For the sake of our argument we perform these symbolic summations by representing a general element $\gamma$ of the free ring by a {\em formal characteristic\,} vector $v(\gamma)$:  this contains the (integer) coefficients of the generators (and there are $|\Sigma|$ formal generators in the left polynomial and $|\Omega|$ formal generators in the right polynomials). To perform the addition of two general elements we add the corresponding characteristic vectors component-wise. Therefore the complexity of such a formal summation is $O(|\Gamma|)$ elementary additions over the integers. Therefore, the overall complexity of our "generalized" Phase 1 would be $O(|\Gamma| \cdot m^{\log_2 3})$ elementary operations over the integers. Fortunately, as will be seen shortly, we can organize this step much more efficiently.

In our  Phase 2, we must compute at each leaf the product of two general integer elements of the free ring. This amounts to computing the pairwise products, each accompanied by the product of the two integer coefficients. At each leaf there are at most $|\Sigma| |\Omega|$ such pairs.

Instead of evaluating these standard products,  we apply a map $\Psi$ from the products of any two generators into the ring $\mathbb{Z}$ of integers. We take, as an example, the case where a fit happens when the text character belong to the subset in the pattern. Then each double product $\sigma \omega$ maps to $1$ iff the text symbol matches to the pattern element and $0$  otherwise. This is exactly Equation (\ref{eq:scalar}), just the scalar product of the corresponding $\bar t$ and $\bar p$ characteristic vectors. Then we extend this map to the product of two general elements in the usual way: the $\Psi$-image of each double product will be accompanied by the product (over the integers) of the two coefficients. (If the fitting conditions are different, then the definition -and the computing method- of the pair-wise product of one element from $\Sigma$ and one from $\Omega$ will differ accordingly.)

In fact, it is easy to see that this is a group homomorphism from the additive group of the free ring to the additive  group of $\mathbb{Z}$.  In this way, each coefficient  in the final product, which is a linear  combination of double products, is mapped into $\mathbb{Z}.$

We remark that by the distributive and commutative laws for integer numbers the formal linear combination of generator elements and the double products of the generators at the bottom are fully interchangeable. Therefore, in Phase I, instead of using formal linear combinations of generating elements, we can perform the standard linear combinations of vectors of dimension $|\Sigma|$ and $|\Omega|$ over $\mathbb{Z}.$ Thus, in each step of Phase I, we have to calculate the linear combination of  two (integer) characteristic vectors of dimension $O(|\Gamma|)$ and store the result as a new integer vector over $\mathbb{Z}.$

In conclusion, instead of introducing the formal characteristic vectors $v(\gamma)$ we just keep the original representations of our text symbols and pattern elements and treat them as integer vectors. Therefore, we need $O(|\Gamma|)$ space to store the current polynomials at each step in Phase I, and the time complexity of Phase I is altogether $O(|\Gamma| m^{\log_2 3}).$

Clearly, a perfect match occurs if and only if the coefficient equals $m.$ On the other hand, if this coefficient is $m' <m$ then  there are exactly $m-m'$ mismatches between $T'$ and $P$.

\bigskip\noindent In Phase 2 of  $KAM$  we perform  $O(m^{\log_2 3})$ pairwise multiplications between elements of the free ring.  In our case the pairwise multiplication is simply the scalar product of two characteristic vectors of dimension $|\Sigma|$, whence our  Phase 2 charges $O\left (|\Sigma| |\Omega| |\Sigma| m^{\log_2 3}\right )$ multiplications over $\mathbb{Z}$ and the same number of additions overall.

\bigskip\noindent In Phase 3 we compute the $\Psi$-image of every addition instead of the additions of general elements of the free ring. In all such steps we have thus just integers as factors. Therefore,  Phase 3 has exactly the same complexity as in the original Karatsuba algorithm, that is, $O(m^{\log_2 3})$.

We can conclude that the overall product of  $T'(x)$ and $  P(x)$  involves no more than  $3r  m^{\log_2 3} + 4 m^{\log_2 3}$ additions and $O(|\Sigma| |\Omega|  m^{\log_2 3})$ pairwise products. Running the algorithm for all consecutive non-overlapping segments of the text  and putting together the resulting product polynomials will complete the procedure, resulting in
\begin{equation}\label{lm:chunk-comp}
O\left ( \frac{n}{m}  |\Sigma|^2 |\Omega| m^{\log_2 3} \right ) ~= ~ O\left ( n |\Sigma|^2 |\Omega| m^{\log_2 3 - 1} \right ) ~= ~ O\left ( {n}|\Sigma|^2 |\Omega| m^{0.585} \right).
\end{equation}
In cases when $|\Omega|$ is not too big (say it remains constant while $n$ becomes longer) then this term is $\eredmeny.$ It is also possible to calculate all possible fitting scores in advance: taking all possible text characters against all possible pattern character classes (including the local numerical values), calculating and storing all occurring values in a corresponding two dimensional array. Then, at the cost of $|\Sigma|^2 |\Omega|$ extra space we can reduce the time complexity from $ O\left ( {n}|\Sigma|^2 |\Omega| m^{0.585} \right)$ to $ O\left ( {n}|\Sigma| m^{0.585}  +|\Sigma|^2 |\Omega| \right) $ which is again only $\eredmeny$.

This concludes the discussion of our claim. \qed

\section{The plug-in procedure}\label{sc:numera}
\noindent
In this section we detail two additional plug-in procedures in order  to exemplify the variety of   incarnations of modulated string searching. Consider first a simple solution for the $\tau$-truncated $L_1$-distance string matching  on character classes in which we assume an infinite value for the constant $b$ (see the detailed description at the beginning of Section \ref{sc:MSS}). We examine the formal multiplications between elements of $\Sigma$ and $\Omega,$ found at the bottom of  the recursion tree. Their representative characteristic  vectors  were called  $t$ and $p$, respectively. Each text element (a characteristic vector) contains one character, while the pattern class may contain several characters. To calculate the "product" of these characteristic vectors, one should find the two characters in the pattern class which are closest to the text character (using, for example, a standard merge), then calculate the smaller $L_1$-distance, and finally truncate it with the constant $\tau.$ Instead we can check the closed $\tau$-ball around the text element to see whether it contains elements of the pattern class and, if the answer is affirmative, perform the necessary calculation. This requires no more than $2\tau + 1$ steps at each multiplication. The subsequent steps are obvious. The solution of the locally bounded $L_1$-distance string matching on character classes is similar.

Next, to demonstrate the method in a more complex context, we compute  the $L_1$-distances at points where pattern and text meet modulated pattern matching conditions such as the above. For this we will need to manage a second characteristic vector pair for text and pattern, respectively, which will store the actual text and pattern elements in form of symbolic linear combinations.

We revisit the multiplication of the largest linear combinations found at the bottom of  the recursion tree. These were called  $t$ and $p$, respectively. They consist of two  characteristic vectors storing $\min (m, |\Sigma|)$ elements  from the text and $\min (m, |\Omega|)$ terms from the pattern.  This time, our multiplication consists of computing the sum of the differences  $(\ell_i - \ell_j)$ that can be formed by taking  one symbol from $t$ and one from $p$.  Consider first pattern elements such that  $\ell_i \geq \ell_j$. Letting $r$ and $r'$ be the number of symbols in $p$ and $t$ respectively, we assume the existence of  index tables $I$ and $I'$ that take  from the value of $\ell_h$ to $h,$ respectively for  $1 \leq h \leq r' $ and $1 \leq h \leq r $. We need  the array $S$ containing at the $h$-th position the value
\begin{equation}\label{eq:distance}
S_{h-1} = \sum_{j=h}^r (\ell_j - \ell_{h-1}) f_j,
\end{equation}
where $f_j$ denotes the multiplicity of run length $\ell_j$.  Clearly,
$$ (\ell_j - \ell_{j-1})(f_r + ... + f_j) = S_{j-1} - S_j$$
so that $S$ can be filled in linear time using
$$S_{j-1} =   S_j  +  (\ell_j - \ell_{j-1})\times F_j $$
where $F_j = f_r + ... + f_j$  is  obtained for all values of $j$ by a single suffix computation on  the frequencies. With the array $S$ in place, the cumulative distance $\Delta$ of a text run length $\ell$ from the pattern runs is computed as follows. Let $\ell_{j-1} \leq \ell < \ell_j$.  Then,
$$
\Delta =   S_j  +  (\ell_j - \ell ) \times F_j .
$$
We deal with the cases $\ell_i \leq \ell_j$ analogously. The overall procedure results in no increase in the time complexity.

It is easy to formulate many additional variants of the problem. For instance, assume that we are still interested in the $L_1$-distance between the pattern and the text at each possible starting position. However, we require now in addition that at each position the text element must fall within a possibly varying, specified neighbourhood of the pattern element. For example, their difference must be never bigger than some {\it a priori\,} assigned value $\tau$ (the previous problem), or it must be always an even number, or, it must be even whenever the difference is at most $h$, odd otherwise. And so on.

\bigskip\noindent
{\bf Acknowledgement}
We are indebted to Amihood Amir for his encouragement and for enlightening discussions.

\bibliographystyle{elsarticle-num}

\end{document}